\documentclass[12pt,aps,prd,superscriptaddress,
nofootinbib,
 amsmath,amssymb,
aps]{revtex4-2}

\usepackage{graphicx}
\usepackage{xcolor}
\usepackage{bm}

\newcommand{\Tv}{T_{\text{vis}}}
\newcommand{\Th}{T_{\text{hid}}}
\newcommand{\trh}{t_{\text{RH}}}
\newcommand{\Mp}{M_{\text{pl}}}

\usepackage{hyperref}
\usepackage{cleveref}

\begin{document}

\preprint{APS/123-QED}

\title{Dark Monopoles, Bounds on Hidden Sectors,\\ and Cosmological Implications}

\author{Donald Liveoak}
\affiliation{%
 Leinweber Institute for Theoretical Physics, Physics Department, University of Michigan, Ann Arbor, MI 48109, USA
}

\author{Anshuman Maharana}
\affiliation{%
 Leinweber Institute for Theoretical Physics, Physics Department, University of Michigan, Ann Arbor, MI 48109, USA
}
\affiliation{Harish-Chandra Research Institute, Chhatnag Road, Jhunsi,
Prayagraj, Uttar Pradesh 211019, India}

\affiliation{Homi Bhabha National Institute, Training School Complex, Anushakti Nagar, Mumbai
400094, India}

\author{James D. Wells}%
\affiliation{%
 Leinweber Institute for Theoretical Physics, Physics Department, University of Michigan, Ann Arbor, MI 48109, USA
}

\date{\today}

\begin{abstract}
{\it Abstract:} Hidden sectors are a generic prediction of string theory compactifications and result in a promising landscape for dark matter model building. We consider the case of hidden sector magnetic monopoles produced via a thermal phase transition in the early Universe and subsequently diluted by pair annihilation. We show that for symmetry-breaking scales $\gtrsim 100\, \text{PeV}$, the monopole abundance is unacceptably high, overclosing the Universe. Our bounds are robust against variations in the initial fraction of energy density deposited in the hidden sector, exhibiting only a weak power-law dependence on this quantity. The bound is substantially tightened in the case of multiple hidden sectors. The standard cosmology may only be recovered if one of the following is true: the hidden sector(s) are non-existent, the hidden sectors have no monopoles with symmetry-breaking scale above 100 PeV, the maximum temperature of each monopole-producing hidden sector after reheating is below its symmetry-breaking scale, or the monopole abundance is diluted by an additional physical mechanism (such as early matter domination, the Langacker-Pi mechanism, or capture by primordial black holes).
\end{abstract}

\maketitle



\section{Introduction}
Dark matter (DM) is one of the most important elements of modern cosmology and astrophysics. Despite this, we still lack an understanding of its microscopic description. There is a zoo of models \cite{bertone2005particle, feng2010dark, cirelli2025darkmatter}, including WIMPs \cite{arcadi2018waning, roszkowski2018wimp}, axions \cite{marsh2016axion, di2020landscape}, WIMPzillas \cite{chung1998superheavy, kolb1999wimpzillas}, and sterile neutrinos \cite{boyarsky2009role}, where DM arises from minimal extensions to the Standard Model (the so-called \textit{visible sector}) of particle physics. 

Equally compelling are models where DM resides in a \textit{hidden sector} with a gauge group independent from the Standard Model. Hidden sector monopoles (or dark monopoles) may be formed if the hidden sector undergoes a thermal phase transition which spontaneously breaks the gauge group $G$ to a subgroup $K$, such that the topology of the vacuum manifold $G/K$ is non-trivial. Topological dark matter has been investigated in \cite{murayama2010topological, gomez2011monopoles}.
Hidden sector monopoles as dark matter have been studied in a variety of contexts \cite{evslin2012dwarf}, including in the presence of vectors and radiation \cite{khoze2014dark, baek2014hidden, brummer2026no} and scenarios of early matter domination \cite{Graesser:2020hiv}.

Importantly, if the initial number density of dark monopoles is too high, and if annihilation is weak, they may overclose the Universe. In this paper, we derive the conditions for overclosure by considering production from a thermal phase transition and a detailed treatment of monopole-antimonopole annihilation. We then discuss the resulting physical implications.

Inflation has emerged as the leading candidate for early Universe cosmology. During the inflationary epoch, all of the energy density of the Universe is in the inflaton field. The decay of the inflaton reheats the Universe and initiates the hot big bang. This picture of a common origin of all constituents of the Universe from the decay of a single field is supported by the absence of isocurvature perturbations \cite{Planck:2018vyg}. 
  
The above implies two possibilities for hidden sectors. The first is where the hidden sector thermalizes with the visible sector (e.g., through kinetic mixing or a Higgs portal). In this case, the hidden sector carries a sizable fraction of the total energy of the Universe. The other possibility is that the hidden sector is effectively isolated and is only able to interact gravitationally with the visible sector. In this case, the two sectors do not reach thermal equilibrium with each other. The energy density of such a hidden sector is determined by the inflaton decay branching ratio $B$. If such a sector thermalizes with itself, it evolves with its own temperature. 

A hidden sector with $B\ll 1$ carries a very small fraction of the total energy density at the end of reheating.  One would expect such sectors to be irrelevant to cosmology. This expectation is borne out for various observables. For example, if the hidden sector has a massless degree of freedom, its leading-order contribution to the dark radiation problem is proportional to $B$ (see  e.g. \cite{Cicoli:2023opf} for a detailed discussion). This implies that hidden sectors with a low branching ratio are irrelevant from the point of view of dark radiation. 

Crucially, we will show that the expectation that small $B$ implies the irrelevance of the hidden sector does not hold in the case of dark monopoles. By carefully modeling the annihilation mechanisms that reduce the monopole abundance, we will find that monopoles with symmetry-breaking scales $\gtrsim 100\, \text{PeV}$ generally overclose the Universe and thus are unacceptable within the standard cosmology. One then would be required to inflate them away by a subsequent period of inflation as originally motivated by Guth \cite{Guth:1980zm}. Additionally, we will show that the final relative energy density of monopoles depends rather weakly on the inflaton branching ratio to the hidden sector ($\sim B^{1/4}$). Thus, it is difficult to avoid overclosure by fine-tuning the model of reheating. This poses strong constraints on the structure of hidden sectors. Furthermore, we will show that in the case of $N$ hidden sectors, the bound is further tightened by a factor $\sim N^{3/4}$. To obtain a present-day energy density less than the observed density of DM, the monopoles must either be diluted by inflation or by an additional physical mechanism (such as early matter domination).

Our results have strong implications for string phenomenology, or any scenario that generically predicts hidden sectors. Hidden sectors arise naturally in string compactifications, and are often
needed for mathematical consistency (see, for example,  heterotic constructions \cite{Giedt:2000bi, Dienes:2007ms, Anderson:2013xka},  intersecting D-brane models \cite{Cvetic:2004ui, Gmeiner:2005vz, Loges:2022mao, Marchesano:2024gul},  F-theory constructions \cite{Taylor:2015ppa},  rational conformal field theory constructions \cite{Dijkstra:2004cc}, and  M-theory constructions \cite{Acharya:2016fge}). The number of hidden sectors is often large. Furthermore, even within a given construction, one can expect hidden sectors with distinct fundamental scales due to warping in the extra dimensions \cite{Hebecker:2006bn}. 
There are a plethora of possibilities for
(cosmologically) stable dark matter candidates in string compactifications. The conserved
quantum number can be associated with discrete/continuous isometries of the geometry, unbroken subgroups of gauge symmetries or charges of non-perturbative objects (see e.g. \cite{Witten:1985xc, Shiu:2003ta, Lebedev:2006kn,
Lebedev:2007hv, Frey:2009qb, Berasaluce-Gonzalez:2011gos,  Halverson:2016nfq, Halverson:2018vbo}). Here, the focus will be on dark monopoles that can arise from such sectors. A single hidden sector that violates our overclosure bounds is disastrous for any model.  We present generic conditions that can lead to evasion of our bounds, including hidden sector fine-tuning, inflation, or post-phase transition dilution.

This paper is structured as follows. In \Cref{sec:theory}, we discuss the class of models that form hidden sector monopoles. In \Cref{sec:production}, we discuss the non-thermal production of monopoles via the Kibble-Zurek mechanism. In \Cref{sec:annihilation}, we present the mechanisms by which hidden sector monopoles may annihilate. We analytically and numerically compute the monopole abundance and overclosure bounds in \Cref{sec:abundance}. In \Cref{sec:discussion}, we discuss the assumptions of our analysis. In \Cref{sec:consequences}, we present the consequences of our findings to cosmology and string phenomenology. We conclude in \Cref{sec:conclusion}.

\section{Theoretical framework}\label{sec:theory}
We first consider the case of a single hidden sector which thermalizes with itself but is thermally isolated from the visible sector. In this case, the temperatures of the hidden and visible sectors will be different. Specifically, we assume that the inflaton branching ratio into the hidden sector is $B$, so that the ratio of the energy densities of the hidden and visible sectors at the time of reheating, $\trh$, is 
\begin{equation}
    \frac{\rho_{\text{hid}}(\trh)}{\rho_{\text{vis}}(\trh)} = \frac{B}{1-B}.
\end{equation}
In the subsequent cosmological evolution, the entropy densities in both the hidden and visible sectors, $s_{\text{hid}}$ and $s_{\text{vis}}$, are conserved. We consider the case in which the energy densities in both sectors are radiation dominated so that
\begin{align}
    s_{\text{vis}} &= \frac{2\pi^2}{45} g_\ast^{\text{vis}}(t) \Tv(t)^3 =\frac{4}{3}\frac{\rho_{\text{vis}}(t)}{\Tv}\\
    s_{\text{hid}}& = \frac{2\pi^2}{45} g_\ast^{\text{hid}}(t) \Th(t)^3 =\frac{4}{3}\frac{\rho_{\text{hid}}(t)}{\Th},
\end{align}
where $\Tv(t)$ is the temperature of the visible sector and $g_\ast^{\text{vis}}(t)$ is the effective number of relativistic degrees of freedom in the visible sector (1 per bosonic spin d.o.f., $7/8$ per fermionic spin d.o.f.). The parameters $\Th(t)$ and $g_\ast^{\text{hid}}(t)$ are defined analogously for the hidden sector. Using the conservation of entropy density, we may derive
\begin{align}
    \frac{\rho_{\text{hid}}(t)}{\rho_{\text{vis}}(t)} &\equiv f(t) =  \left( \frac{g_\ast^{\rm hid}(t_{\rm rh})}{
 g_\ast^{\rm vis}(t_{\rm rh})} \right)^{1/3}
 \left( \frac{g_\ast^{\rm vis}(t)}{
 g_\ast^{\rm hid}(t)} \right)^{1/3}  \left( \frac{B}{1-B} \right), \\
  \frac{\Th(t)}{\Tv(t)} &\equiv \eta(t) =\left( \frac{g_\ast^{\rm hid}(t_{\rm rh})}{
 g_\ast^{\rm vis}(t_{\rm rh})} \right)^{1/12}
 \left( \frac{g_\ast^{\rm vis}(t)}{
 g_\ast^{\rm hid}(t)} \right)^{1/3}  \left( \frac{B}{1-B} \right)^{1/4}.\label{ratio}
 \end{align}
The Hubble parameter is 
\begin{equation}
    H = \sqrt{\frac{\rho_{\text{vis}}+\rho_{\text{hid}}}{3\Mp^2}}=\sqrt{\frac{\rho_{\text{vis}}(1+f(t))}{3\Mp^2}},
\end{equation}
where $\Mp\approx2.4\times10^{18}\,\text{GeV}$ is the reduced Planck mass.

To evade observational bounds on dark radiation, we now specialize to the limit $B\ll1$. For $B=O(1)$, the energy density of the hidden sector contributes to a substantial effective number of relativistic degrees of freedom at BBN, which is strongly constrained by Planck observations \cite{collaboration2020planck}. For simplicity, we assume that the effective number of relativistic degrees of freedom in each sector is constant, $g_\ast^{\text{vis}}=g_\ast^{\text{vis}}(\trh)$ and $g_\ast^{\text{hid}}=g_\ast^{\text{hid}}(\trh)$. We note that our analysis may be easily generalized to the case where $g_\ast^{\text{hid}}$ and $g_\ast^{\text{vis}}$ are time-varying. Additionally, we neglect the dependence of $f$ and $\eta$ on $g_\ast^{\text{hid}}$ and $g_\ast^{\text{vis}}$ by approximating $g_\ast^{\text{hid}}/g_\ast^{\text{vis}} \approx 1$. We will later restore the appropriate factors of $g_\ast^{\text{hid}}$ and $g_\ast^{\text{vis}}$ by making the replacement $B^{1/4} \to B^{1/4}(g_\ast^\text{vis}/g_\ast^\text{hid})^{1/4}$, as suggested by \Cref{ratio}.

These assumptions imply $f(t)\approx B \ll1$ and $\eta(t) \approx B^{1/4}$, and the cosmological evolution is dominated by the energy density in the visible sector. Then, the Hubble parameter reduces to the usual form $H=\Tv^2/C\Mp$ where $C=\sqrt{90/\pi^2 g_\ast^{\text{vis}}}$.

We assume that at some critical hidden sector temperature $T_c$, the gauge group of the hidden sector is spontaneously broken from $G \to K$. This phase transition corresponds to a visible sector temperature $\Tv = T_c B^{-1/4}$. If the second homotopy group of the manifold of degenerate vacua $\pi_2(G/K)$ is non-trivial, there will be topologically stable 't Hooft-Polyakov magnetic monopoles of mass $m_M \sim 2\pi \hat{v}/e$, where $\hat{v}\sim T_c$ is the vacuum expectation value of the dark Higgs and $e$ is the hidden sector electric charge \cite{t1974magnetic, polyakov1974particle}. At the phase transition, monopoles will be copiously produced according to the Kibble-Zurek mechanism, which we will describe in \Cref{sec:production}. The minimal magnetic charge of the monopoles is $h=2\pi/e$ \footnote{Dirac quantization requires $h=2k\pi/e$ for some integer $k$ which depends on the global topology of the hidden sector gauge group; here, we choose $k=1$, but our analysis is easily extended to $k>1$.}.

After the phase transition, the evolution of the number density of monopoles $n_M$ is governed by the Boltzmann equation
\begin{equation}
    \dot{n}_M=-3Hn_M -D\left(n_M^2-({n_M^{\text{eq}}}(\Th))^2\right),
\end{equation}
where $D$ is the thermally-averaged annihilation rate coefficient and encodes the mechanism(s) for monopole-antimonopole annihilation, and $n_M^{\text{eq}}(\Th)$ is the number density of monopoles at thermal equilibrium. We will see that in the case of Kibble-Zurek production of monopoles at the phase transition, $n_M\gg n_M^{\text{eq}}$ at all stages of cosmological evolution, and thus the thermal production of monopoles may be neglected.

\section{Monopole production}\label{sec:production}
We assume that the hidden sector is initially at a temperature $\Th > T_c$, and monopoles are produced as the result of a second-order phase transition (SOPT) at $\Th = T_c$. We will closely follow the presentation of \cite{murayama2010topological}.

A bound on the number density of monopoles produced in this scenario was first provided by Kibble \cite{kibble1976topology}. Since the correlation length in a SOPT diverges but is limited by the causal horizon $d \sim H^{-1}$, we must have a minimal number density of monopoles of approximately one per Hubble volume $n_M H^{-3} \sim 1$. Zurek \cite{zurek1985cosmological} later refined this conservative estimate, accounting for the freeze-out due to finite quenching time, demonstrating that the true number density of monopoles is typically several orders of magnitude larger than one per horizon. We will now present the key steps of the so-called Kibble-Zurek mechanism.

We parameterize the correlation length $\xi$ and relaxation time $\tau$ of the phase transition via
\begin{align}
    \xi &= \xi_0 |\epsilon|^{-\nu} & \tau = \tau_0|\epsilon|^{-\mu},
\end{align}
where $\epsilon\equiv (T_c-T)/T_c$, $\nu$ and $\mu$ are critical exponents, and $\xi_0$ and $\tau_0$ are the initial correlation length and relaxation time, respectively.
As the Universe cools, we will eventually have $\Th \sim T_c$ at some time $t=t_c$. Near this point, we may expand $\epsilon$ to leading order as
\begin{equation}
    \epsilon \approx \epsilon(\Th=T_c)+\frac{d\epsilon}{dt}\Bigg|_{\Th=T_c} (t-t_c)= \frac{|d\Th/dt|}{T_c}(t-t_c).
\end{equation}
Then, at leading order, the characteristic timescale of quenching is 
\begin{equation}
    \tau_Q \equiv \frac{T_c}{|d\Th/dt|} \approx \frac{(t-t_c)}{\epsilon}.
\end{equation}
During radiation domination, $\tau_Q(\Th=T_c)=2t_c=H(\Th=T_c)^{-1}$. At some critical time $t_\ast$, we will have $|t_\ast - t_c| < \tau$; at this point, the hidden sector cannot maintain thermal equilibrium, and the correlation length is frozen out and will maintain its value until after the phase transition. We solve to find $|\epsilon(t_\ast)|= (\tau_Q/\tau_0)^{-1/(1+\mu)}$ and thus the correlation length is frozen at 
\begin{equation}
    \xi(t_\ast) = \xi_0\left(\frac{\tau_Q}{\tau_0}\right)^{\nu/(1+\mu)}.
\end{equation}

Assuming a Landau-Ginzburg Hamiltonian with potential $V(\phi)=(\Th-T_c)M \phi^2+\frac{1}{2}\lambda \phi^4$ with $M$ the mass parameter and $\lambda$ the self-coupling of the dark Higgs field $\phi$, the critical exponents are classically $\mu=\nu=1/2$ \cite{murayama2010topological}. Additionally, $M = K\lambda T_c$ where $K=\mathcal{O}(1)$ is a model-dependent factor. The zero-temperature dark Higgs vev is $\hat{v} = (M T_c/\lambda)^{1/2} = K^{1/2}T_c$. For the rest of our analysis, we will take $K=1$ so that $\hat{v} = T_c$. \footnote{One may easily restore the appropriate factors of $K$, given a specific hidden sector gauge group, dark Higgs representation, and symmetry-breaking potential.}

We assume an initial correlation length and relaxation time $\xi_0 = \tau_0 = 1/{\sqrt{\lambda} T_c}$, so the correlation length is
\begin{equation}
    \xi(t_\ast) = \left(\frac{\sqrt{\lambda} T_c}{H(\Th=T_c)}\right)^{1/3} \frac{1}{\sqrt{\lambda} T_c} = \left(\lambda H(\Th=T_c) T_c^2\right)^{-1/3}.
\end{equation}
We expect one monopole per correlation volume $\xi(t_\ast)
^3$, implying an initial monopole density
\begin{equation}
    n_M \sim \xi(t_\ast)^{-3} \sim \lambda H(\Th=T_c)T_c^2.
\end{equation}

We note that $\xi(t_\ast)H\sim(T_c/\Mp)^{2/3}$, so for phase transitions at temperatures at or below the GUT scale $\sim 10^{16} \, \text{GeV}$, there are a large number of monopoles per causal horizon $n_M H^{-3} \sim (\Mp/T_c)^2 \gg 1$.

\section{Annihilation mechanisms} \label{sec:annihilation}
When produced at high abundances, the monopole density will decrease via the formation of monopole-antimonopole Coulomb bound states which subsequently cascade and annihilate \cite{zeldovich1978concentration, preskill1979cosmological}. This process occurs when the distance between a monopole and antimonopole is less than the Coulomb capture radius $a_c=h^2/4\pi\Th$, determined by equating the Coulomb potential energy $h^2/4\pi a_c$ to the typical kinetic energy $\sim\Th$. Schematically, the annihilation process is
\begin{equation}
    M+\bar{M}\to M\bar{M}\to \gamma'\gamma',\phi\phi, \dots
\end{equation}
where $\gamma'$ is the dark photon. Importantly, though the cosmological evolution is dominated by the visible sector temperature $\Tv$, the monopole annihilation dynamics only depend on the temperature of the hidden sector $\Th$.

At high temperatures, the most efficient mechanism for monopole annihilation is diffusive capture via scattering with a thermal bath of relativistic fermions charged under the hidden sector gauge group. The thermally averaged cross section for scattering of a magnetic monopole from a particle of charge $q_F$ is $\sigma_{MF\to MF}\sim(hq_F/4\pi)^2 \Th^{-2}$ \cite{preskill1979cosmological}. Thus, for a single relativistic fermion species $F$ with equilibrium number density $n_F(\Th)=(3/4\pi^2)g_F\zeta(3) \Th^3$, where $\zeta(z)$ is the Riemann zeta function and $g_F$ is the number of fermion spin states, the rate of monopole-fermion scattering events is
\begin{equation}
    \Gamma_{MF\to MF}=\sigma_{MF\to MF} \,n_F(\Th)=\frac{3}{4\pi^2}\zeta(3) \left(\frac{hq_F}{4\pi}\right)^2 \Th g_F.
\end{equation}
Large-angle scattering occurs once per $\sim m_M/\Th$ scattering events \cite{preskill1979cosmological}. Summing over the contributions from all fermion degrees of freedom, the total rate of large-angle monopole-fermion scattering is
\begin{equation}\label{eq:bdef}
    \Gamma_{\text{large-angle}} = \frac{\Th}{m_M}\frac{3}{4\pi^2} \zeta(3) \Th\sum_{F} g_F\left(\frac{hq_F}{4\pi}\right)^2 \equiv b\Th^2/m_M.
\end{equation}
The quantity $b$ parametrizes the number of relativistic fermion degrees of freedom in the hidden sector plasma. For our analysis, we choose a fiducial value of $b\approx 10$, consistent with \cite{preskill1979cosmological}. The typical time between large-angle scattering events is $\tau_{\text{large-angle}} = m_Mb^{-1}\Th^{-2}$. 

The effective rate of monopole annihilation is determined by the flux of monopoles through a sphere of radius $a_c$ around an antimonopole. As the monopole approaches the antimonopole, it acquires a drift velocity
\begin{equation}
    v_{M} \sim \frac{h^2}{4\pi a_c^2} \frac{\tau_{\text{large-angle}}}{m_M}
\end{equation}
and thus the overall rate of monopole annihilation is 
\begin{equation}
    \Gamma = 4\pi a_c^2 n_M v_{M} \sim n_M\frac{h^2\tau_{\text{large-angle}}}{m_M}\sim n_M\frac{h^2}{b \Th^2},
\end{equation}
so the annihilation rate coefficient $D \equiv \Gamma n_M^{-1}=h^2 b^{-1} \Th^{-2}$. This process persists until the mean free path of the monopoles, $\ell \sim (\Th/m_M)^{1/2} \tau_{\text{large-angle}}$ is equal to $a_c$; this implies that annihilation remains efficient until $\Th=\Th^{\text{stop}}=m_M b^{-2}(4\pi/h^2)^2$.

If the hidden sector does not contain light fermions and $\Th$ is greater than $m_{X'}=e\hat{v}$, the thermal bath will contain stable relativistic gauge bosons $X'$. The contributions of monopole scattering off of the $X'$ bosons are analogous to those of hidden sector fermions, and may be accounted for by including their spin degrees of freedom in the definition of $b$. However, once $\Th \sim m_{X'}$, the abundance of thermal $X'$ bosons is exponentially diluted, and thus this annihilation mechanism only remains efficient until $\Th=\Th^{\text{stop}}=e\hat{v}$ \cite{Khoze:2014woa}.

As the Universe cools, diffusive capture will eventually become inefficient. At this point, monopoles and antimonopoles may only be captured by emission of dark radiation (bremsstrahlung). This results in an annihilation rate coefficient of $D\sim(h^2/4\pi)^2 m_M^{-2} (m_M/\Th)^{9/10}$ \cite{preskill1979cosmological}. The annihilation rate due to dark bremsstrahlung is eventually cut off by the expansion of the Universe, and the monopole abundance is frozen out. We will demonstrate this explicitly in \Cref{sec:abundance}. 

\section{Monopole abundance} \label{sec:abundance}

\subsection{Analytical results}
We first estimate the monopole abundance today assuming a thermal bath of dark fermions which catalyze the capture and annihilation of monopoles. We integrate the Boltzmann equation from $\Tv=\Tv^i\equiv T_c B^{-1/4}$ until fermion-catalyzed annihilation ceases at 
\begin{equation}\label{eq:Tstopvis}
    \Tv^{\text{stop}}= m_M b^{-2} (4\pi/h^2)^2 B^{-1/4}.
\end{equation}
After $\Tv=\Tv^{\text{stop}}$, the quantity $n_M/s\approx n_M/s_{\text{vis}}\sim n_M/\Tv^3$ is conserved during the subsequent evolution.

In the case of non-thermal Kibble-Zurek production of monopoles, the initial abundance is much greater than the equilibrium abundance
\begin{equation}
    \frac{n_M^i}{(\Tv^i)^3} = \frac{\lambda B^{1/4} T_c}{C\Mp} \gg \frac{n_M^{\text{eq}}(\Tv^i)}{(\Tv^i)^3} \sim \frac{B^{1/4}}{2\pi^2} \frac{m_M^2}{(\Tv^i)^2} K_2(B^{-1/4}m_M/\Tv^i),
\end{equation}
where $K_2(x)$ is the modified Bessel function of the second kind. Since $K_2(x)\sim x^{-1/2}e^{-x}$, the equilibrium abundance is always small compared to the non-thermal abundance, and thus the term in the Boltzmann equation $\propto (n_M^{\text{eq}})^2$ may be neglected for the entirety of the evolution.

For a power-law annihilation term $D=(A/m_M^2) (m_M/\Th)^2=(AB^{-1/2}/m_M^2)(m_M/\Tv)^2$, the Boltzmann equation may then be integrated to find \cite{preskill1979cosmological}
\begin{equation}
    \frac{\Tv^3}{n_M(\Tv)} = \frac{C\Mp B^{-1/4}}{\lambda T_c} + \frac{AC \Mp B^{-1/2}}{m_M}\left(\frac{m_M}{\Tv}-\frac{m_M}{\Tv^i}\right). 
\end{equation}
For fermion-catalyzed annihilation, the dominant term at $\Tv=\Tv^{\text{stop}}$ is
\begin{equation}
    \left(\frac{n_M(\Tv)}{\Tv^3}\right)_{\Tv=\Tv^{\text{stop}}} = \frac{\Tv^{\text{stop}}}{AC\Mp}\sqrt{B}= \frac{16\pi^2}{bh^6} \frac{m_M}{C\Mp}B^{1/4}.
\end{equation}
Since $s_{\text{vis}} = (2\pi^2/45)g_\ast^{\text{vis}}\Tv^3$ we may write the final comoving abundance as
\begin{equation}
    Y^\infty_M\equiv \left(\frac{n_M}{s}\right)_{\Tv=\Tv^{\text{stop}}} \approx \frac{360}{bh^6g_\ast^{\text{vis}}} \frac{m_M}{C\Mp} B^{1/4}.
\end{equation}
The fractional contribution to today's energy density is 
\begin{align}
\label{omono}
 \Omega_{M} = \frac{ Y^\infty_M s_0 m_M}{3 H^2 _0 \Mp^2} &= \frac{43}{33}\left(\frac{\pi^{2}}{90\left(g_{*}^{\text{vis}}g_{*}^{\text{hid}}\right)^{1/2}}\right)^{1/2}\left(\frac{e^{2}}{\pi}\right)^{2}\frac{1}{b}\left(\frac{T_0^{3}}{M_{\text{pl}}H_0^2}\right)B^{1/4}\left(\frac{\hat{v}}{M_{\text{pl}}}\right)^{2}\\
 &=0.382 \times \left(\frac{\hat{v}}{100\, \text{PeV}}\right)^2\left(\frac{B}{10^{-3}}\right)^{1/4}\left(\frac{b}{10}\right)^{-1}\left(\frac{e}{0.2}\right)^4 \left(\frac{g_\ast^{\text{vis}}}{200}\right)^{-1/4}\left(\frac{g_\ast^{\text{hid}}}{200}\right)^{-1/4},
\end{align}
where $H_0$ is the Hubble constant and $s_0=(2\pi^2/45)(43/11)T_0^3$ is the present-day entropy density, with $T_0$ the CMB temperature. Note that we restored the factor of $g_\ast^{\text{hid}}$ via the replacement $B^{1/4} \to B^{1/4} (g_\ast^\text{vis}/g_\ast^{\text{hid}})^{1/4}$ as suggested by \Cref{ratio}. In terms of the symmetry-breaking scale $\hat{v}$,
\begin{equation}\label{vformu}
  \frac{\hat{v}}{100\, \text{PeV}} = 
   \left(\frac{\Omega_{M}}{0.382} \right)^{1/2} 
   \left(\frac{B}{10^{-3}} \right)^{-1/8}
  \left({\frac{e}{0.2}} \right)^{-2} \left({\frac{b}{10}}\right)^{1/2} \left(\frac{g_\ast^{\text{vis}}}{200}\right)^{1/8}\left(\frac{g_\ast^{\text{hid}}}{200}\right)^{1/8}.
  \end{equation}
We see that for symmetry-breaking scales $\hat{v} \gtrsim 100\, \text{PeV}$, the Universe is overclosed by hidden sector monopoles ($\Omega_M \sim 1$). Since $\Omega_M \sim B^{1/4}$, this problem may not be easily resolved by fine-tuning $B$ to be small. For instance, choosing $B=10^{-9}$ and all other parameters at their fiducial values, the bound becomes $\hat{v} \lesssim 560\, \text{PeV}$.

\subsection{Numerical results}
Now, we numerically integrate the Boltzmann equation to account for the effects of several annihilation mechanisms simultaneously. It is convenient to write the Boltzmann equation in the form
\begin{equation}
    \frac{dY_M}{dx}= -\frac{Ds}{Hx}Y_M(x)^2 \approx -\frac{Ds_{\text{vis}}}{Hx} Y_M(x)^2,
\end{equation}
where $Y_M\equiv n_M/s\approx n_M/s_{\text{vis}}$ and $x\equiv m_M/\Tv$. As before, we integrate from $\Tv=T_c B^{-1/4}$ until fermion-catalyzed annihilation becomes inefficient (\Cref{eq:Tstopvis}). We specify three forms for the annihilation term:
\begin{align}
    D_{\text{rad}}&=\left(\frac{h^2}{4\pi}\right)^2 m_M^{-2} B^{-9/40} x^{9/10}\\
    D_{\text{rad},f}&= \left(\frac{h^2}{4\pi}\right)^2 m_M^{-2} B^{-9/40} x^{9/10} + \frac{h^2 m_M^{-2} B^{-1/2} x^2}{b_f}\theta(\Th-m_Mb_f^{-2}(4\pi/h^2)^2)\\
    D_{\text{rad}, X'} &=\left(\frac{h^2}{4\pi}\right)^2 m_M^{-2} B^{-9/40} x^{9/10} + \frac{h^2 m_M^{-2} B^{-1/2} x^2}{b_{X'}}\theta(\Th-e\hat{v})
\end{align}
where $b_f\approx 10$ and $b_{X'}\approx 0.55$ are defined by \Cref{eq:bdef}, summing over the spin states of the fermions and gauge bosons, respectively. \footnote{Here, we assume that the hidden sector spontaneous symmetry-breaking pattern is $SU(2) \to U(1)$ so that there are two gauge bosons charged under the hidden $U(1)$, each with three polarizations. Our result is easily generalizable to higher rank gauge groups, which result in a larger $b_{X'}$.} Each term includes the contribution to annihilation via the emission of dark radiation; $D_{\text{rad},f}$ and $D_{\text{rad}, X'}$ include fermion and $X'$ boson catalyzed annihilation, respectively.

We integrate the Boltzmann equation using an implicit Runge-Kutta integrator of order $5$ to maintain numerical precision despite $Y_M$ ranging several orders of magnitude. Additionally, we fix $\lambda=0.1$ and $g_\ast^{\text{vis}}=g_\ast^{\text{hid}}=106.75$, though we note that our results do not depend sensitively on this choice. As expected, we find that for the annihilation terms above, the comoving abundance $Y_M$ approaches a constant value for large $x$.

\begin{figure}
    \centering
    \includegraphics[width=\linewidth]{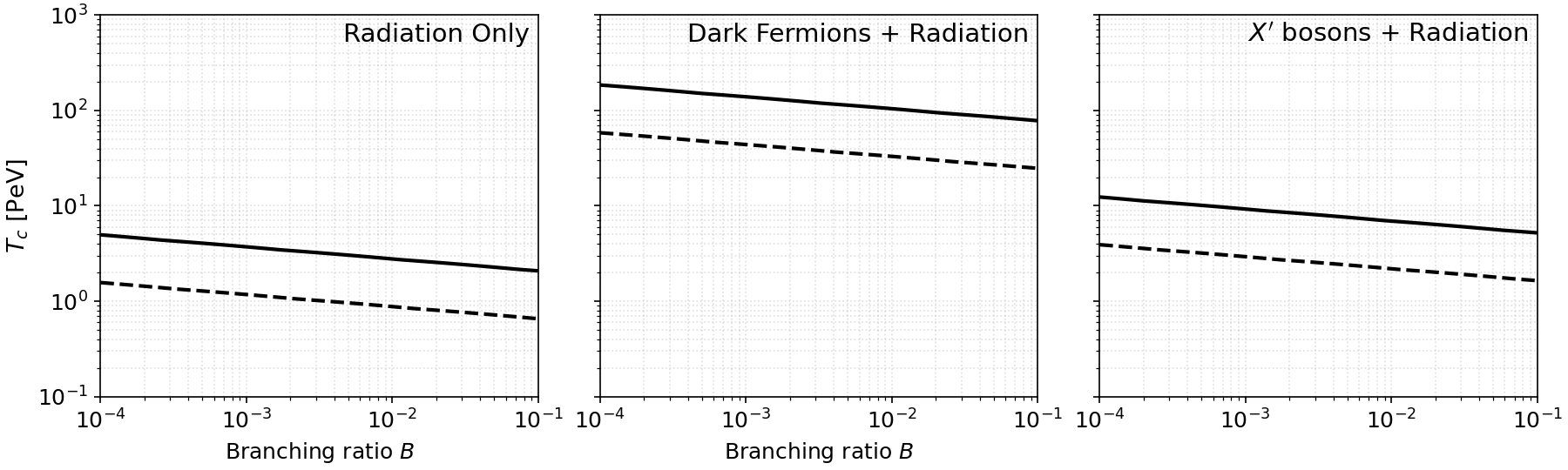}
    \caption{The present-day relative energy density of dark monopoles, $\Omega_M$, as a function of $T_c$ (critical temperature) and $B$ (inflaton branching fraction to the hidden sector). The solid line depicts $\Omega_M =1$, and the dashed line depicts $\Omega_M=0.1$.}
    \label{fig:B-Tc}
\end{figure}

The boundary $\Omega_M=1$ as a function of $T_c$ and $B$ is shown in \Cref{fig:B-Tc}, assuming $e=0.2$. We see that to avoid overclosure, we must have $T_c \lesssim 100 \, \text{PeV}$ if there are dark fermions, and $T_c \lesssim 10 \, \text{PeV}$ if there are no dark fermions. As expected, $\Omega_M$ is a slowly varying function of $B$; to avoid overclosure at $T_c \gtrsim 10-100\, \text{PeV}$, $B$ must be tuned to be extremely small.

\begin{figure*}
    \centering
    \includegraphics[width=\linewidth]{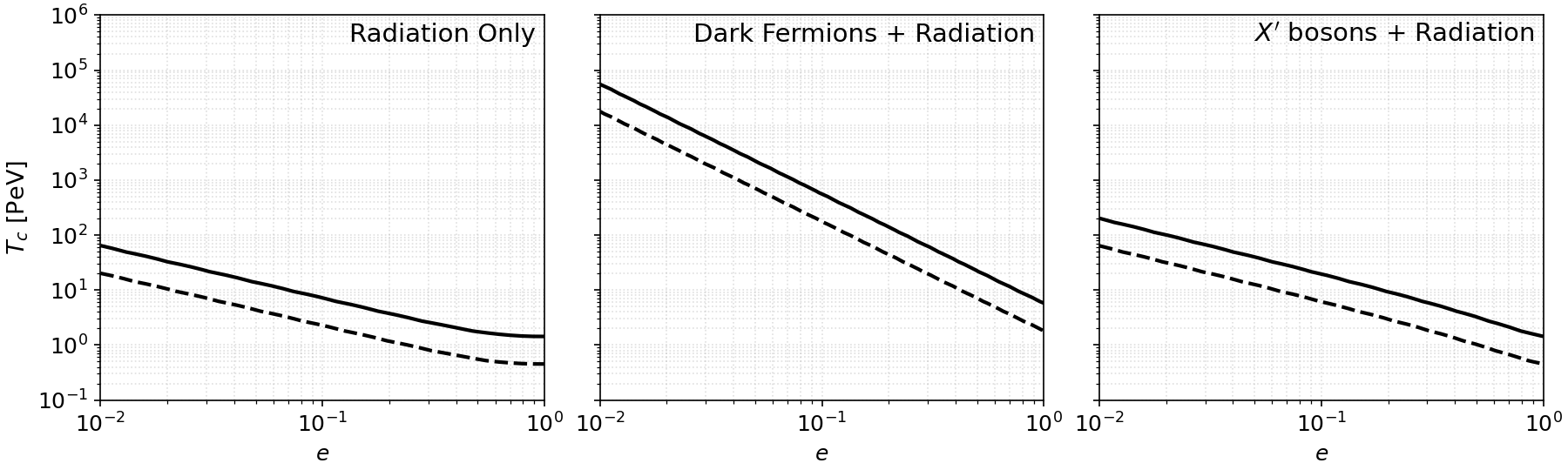}
    \caption{The present-day relative energy density of dark monopoles, $\Omega_M$, as a function of $T_c$ (critical temperature) and $e$ (hidden sector electric charge). The solid line depicts $\Omega_M =1$, and the dashed line depicts $\Omega_M=0.1$.}
    \label{fig:g-Tc}
\end{figure*}

\Cref{fig:g-Tc} depicts the overclosure bound for $B=10^{-3}$ as a function of $T_c$ and dark electric charge $e$. For fermion-catalyzed monopole annihilation, the bound depends strongly on $e$, since the annihilation term $D\propto h^2 \sim e^{-2}$ and annihilation remains efficient until $\Tv^{\text{stop}}\propto 1/h^4 \sim e^4$. For $e \ll 1$, fermion-catalyzed monopole annihilation is more efficient and persists until substantially lower temperatures. Thus, overclosure may be avoided at $T_c \gg 100\, \text{PeV}$ if the electric charge is tuned to be small. We will discuss this in \Cref{sec:discussion}.

\subsection{Multiple hidden sectors}
We now turn to the scenario of multiple hidden sectors. Consider the case of $N$ non-interacting hidden sectors which each have symmetry-breaking pattern $G_i \to K_i$, such that $\pi_2(G_i/K_i)$ is non-trivial. Each sector will admit topologically stable monopole solutions. As before, we assume $(1-\mathcal{B})$ is the proportion of energy density deposited in the visible sector during reheating. Then, as before, we have
\begin{equation}
    B_i^{1/4} \sim \frac{T_i}{T_{\text{vis}}},
\end{equation}
where $B_i$ is the branching ratio of the inflaton decaying into the $i^{\text{th}}$ hidden sector, and $T_i$ is the temperature of the $i^{\text{th}}$ hidden sector. We assume that the total branching ratio into the hidden sectors $\mathcal{B}=\sum_i B_i \ll 1$, so that the cosmological evolution is dominated by the visible sector.

The calculation proceeds identically as before, with each sector acquiring an initial non-thermal monopole density via the Kibble-Zurek mechanism at its respective critical temperature. We specialize to the case in which each sector has identical symmetry-breaking scale $\hat{v}\sim T_c$, Higgs self-coupling $\lambda$, and electric charge $e$. Then, the density of each type of monopole produced at the phase transition is $n_M^i\sim H\lambda T_c^2$, as before. We can then integrate $N$ copies of the Boltzmann equation from the critical temperature until the present epoch. In each formula, we make the replacement $B \to B_i = \mathcal{B}/N$, and multiply the final fractional energy density by $N$. This results in a total fractional energy density
\begin{equation}
    \Omega_{M,\text{total}} = \sum_{i} \Omega_{M,i} \propto N\left(\mathcal{B}
    /N\right)^{1/4} = \mathcal{B}^{1/4} N^{3/4},
\end{equation}
where $\Omega_{M,i}$ is the fractional energy density of monopoles in the $i^{\text{th}}$ hidden sector. Thus, the overclosure bound is significantly tightened in the case of a large number of identical hidden sectors.


\section{Discussion}\label{sec:discussion}
In our analysis, we assumed that the hidden sector(s) were thermally isolated from the visible sector. In reality, the visible and hidden sectors may exchange energy via a Higgs portal interaction term of the form $\lambda_{\Phi\phi } \phi^2 |\Phi|^2$, with $\Phi$ the Standard Model Higgs field. A priori, there is no reason for $\lambda_{\Phi \phi}$ to be suppressed, so it may be the case that the visible and hidden sectors fully or partially thermalize with one another. 

In the case that the thermalization time implied by the Higgs portal interaction $\tau_h$ is small compared to $H^{-1}$, the two sectors will rapidly thermalize and reach a common temperature $\Th = \Tv$. In this case, our analysis applies with $B\sim 1$ up to $O(1)$ factors accounting for the entropy distribution between the two sectors. If $\tau_h \gtrsim H^{-1}$ but not by much, the hidden sector may partially thermalize with the visible sector and be heated to $\Th=\gamma \Tv$, with $\gamma > B^{1/4}$. Again, in this case, our analysis holds with the substitution $B^{1/4} \to \gamma$, up to $O(1)$ factors due to nontrivial entropy density in the hidden sector.

Additionally, we assumed that the hidden sector phase transition which produces monopoles was second-order. This assumption depends on the relative order of magnitude of the hidden sector electric charge $e$ and the dark Higgs self-coupling $\lambda$. For sufficiently small $\lambda$, the phase transition may be first-order \cite{brummer2026no}. Likewise, a first-order phase transition may develop if higher-order operators play a significant role \cite{Grojean:2004xa}. In the case of first-order phase transitions, the initial monopole abundance is determined by the average bubble radius at percolation $R_p$. A detailed treatment of hidden sector monopoles in first-order phase transitions is available in \cite{brummer2026no} for the case that the hidden and visible sectors are thermalized and no light hidden sector fermions are present. In certain regions of $(e,\lambda)$ parameter space, the initial abundance of dark monopoles may be sufficiently small, even with $\hat{v} \gg \, \text{100 PeV}$.

We now discuss the extent to which the hidden sector energy density may be fine-tuned to avoid monopole overclosure. To gain an understanding of this, we must consider global models which include both the Standard Model and an inflationary sector. Given the UV sensitivity of inflation, these models should be analyzed in a UV-complete setting such as string theory. In the case of a supersymmetric theory, superpartner scales of PeV or lower may provide a potentially safe upper bound for hidden sector dynamics and symmetry breaking.
 
The program of constructing such models in string theory is in its infancy \cite{Cicoli:2017shd}. Here, we provide a quick estimate of the suppression that can arise from geometric separation of degrees of freedom. Consider a setting where the Standard Model arises from branes wrapping a local geometric cycle, with the modulus corresponding to the same geometric cycle playing the role of the inflaton. One expects inflaton couplings to be maximally suppressed for hidden sectors that arise from branes wrapping other local cycles which are geometrically separated from the modulus corresponding to the inflaton. For the case of blow-up moduli, the relative strength of the inflaton coupling to a hidden sector and the visible sector scales as the inverse volume of compactification in string units $(\mathcal{V}^{-1})$ \cite{Acharya:2018deu}. Therefore, the relative decay rate (and thus the branching ratio) of the inflaton into the hidden sector scales as $B\sim \mathcal{V}^{-2}$. Thus, for $N$ hidden sectors, $N^{3/4} B^{1/4} \sim N^{3/4} \mathcal{V}^{-1/2}$, implying that one requires large volumes for any significant suppression of the hidden sector energy density. 

The abundance of dark monopoles may be reduced during a period of early matter domination, as explored in \cite{Graesser:2020hiv} (see  \cite{Acharya:2017szw} for a similar discussion in the context of the dark glueball problem). In this scenario, one or more sectors enter matter domination before the hidden sector containing the monopoles can dominate the energy density of the Universe. If these decay predominantly to the visible sector, there can be significant entropy injection and dilution of the dark monopole abundance. In a similar vein, entropy injection from a first-order phase transition may also effectively dilute the monopole abundance \cite{hambye2018super}.

The dark monopole problem may also be solved by invoking other non-inflationary solutions to the GUT monopole problem. One such solution is the Langacker-Pi mechanism, in which the $U(1)$ symmetry associated with the magnetic flux of the dark monopoles is subsequently broken as the hidden sector cools. Then, the flux of the dark monopoles is confined into flux tubes (cosmic strings) which connect dark monopoles and anti-monopoles and facilitate efficient annihilation \cite{langacker1980magnetic}. This mechanism may also lead to an observable gravitational wave signal if the flux tubes accelerate the monopoles to relativistic speeds \cite{PhysRevD.106.075030, PhysRevD.110.016001}. Additionally, the monopole problem may be avoided via enhanced monopole annihilation due to Weyl conformal symmetry breaking \cite{perri2026noninflationary} and monopole capture by primordial black holes \cite{Stojkovic:2004hz}.

Finally, we note that our bounds may be evaded for approximately global monopoles ($e\ll 1$). In this scenario, the magnetic charge $h$ is substantially enhanced, thus increasing the efficiency of monopole-antimonopole capture and annihilation. However, we emphasize that in order to avoid overclosure for $T_c\gtrsim 100\, \text{PeV}$ via fine-tuning $e\to 0$, we must have small electric charges in all hidden sectors; if a single sector has $e \gtrsim 0.2$, overclosure cannot be avoided at this spontaneous symmetry-breaking scale. Additionally, if a hidden sector has no charged fermions, $e \gtrsim 0.01$ leads to overclosure.

\section{Physical consequences}\label{sec:consequences}

We have demonstrated that for a generic hidden sector which permits monopole solutions as a result of a spontaneously broken gauge symmetry, monopoles overclose the Universe unless $T_c \lesssim 100 \, \text{PeV}$. This has strong consequences for the cosmological history of the Universe. Assuming an ultraviolet theory that produces a large number of hidden sectors, one of the following must be true in order to preserve the standard cosmology.

\begin{enumerate}
    \item The hidden sectors are severely constrained. Specifically, no hidden sector may have a spontaneous symmetry-breaking scale $\gtrsim 100\, \text{PeV}$, unless that sector is very weakly coupled ($e \ll 0.1$) and has a large spectrum of charged fermions, or has trivial second homotopy group $\pi_2(G/K)$.
    \item The density of hidden sector monopoles is diluted during inflation. In this case, the reheating temperature of each hidden sector must be below its respective spontaneous symmetry-breaking scale to prevent symmetry restoration and the subsequent formation of monopoles via the Kibble-Zurek mechanism.
    \item Dark monopooles are diluted via a non-inflationary physical mechanism. Such mechanisms include early matter domination, entropy injection in a first-order phase transition, the Langacker-Pi mechanism, Weyl conformal symmetry breaking, and capture by primordial black holes.
\end{enumerate}

\section{Conclusion}\label{sec:conclusion}
In this paper, we considered the cosmological evolution of hidden sector monopoles, carefully accounting for their non-thermal production and annihilation. We showed that the fractional energy density of monopoles scales as $\Omega_M \sim B^{1/4}$, where $B$ is the branching ratio of the inflaton into the hidden sector. For symmetry-breaking scales $\gtrsim 100\, \text{PeV}$, we have shown that the monopoles overclose the Universe ($\Omega_M \gtrsim 1$). In the case of $N$ identical hidden sectors, the overclosure bound is significantly tightened ($\Omega_{M,\text{total}} \sim B^{1/4}N^{3/4}$). To obtain an acceptably low present-day energy density of dark monopoles, we must have at least one of the following: the maximum temperature of each sector after reheating is below its spontaneous symmetry-breaking scale, the hidden sectors are severely fine-tuned ($e\ll 0.1$ for all sectors), or monopoles are diluted during a period of early matter domination.

Future studies should consider how our bounds change for monopoles produced in a more general class of thermal histories, including first-order phase transitions. Additionally, it would be interesting to carefully model total or partial thermalization between the hidden and visible sectors through a Higgs portal. Finally, it remains compelling to explore our bounds in explicit string compactifications with multiple hidden sectors.

\begin{acknowledgments}
DL is supported by the National Science Foundation Graduate Research Fellowship Program. AM would like to thank the Leinweber Institute for Theoretical Physics for supporting his sabbatical visit at the University of Michigan, Ann Arbor. JW acknowledges support from the Leinweber Foundation.
\end{acknowledgments}


\bibliographystyle{apsrev4-2}
\bibliography{biblio}

\end{document}